\documentclass[graybox]{./styles/svmult}

\usepackage{mathptmx}       % selects Times Roman as basic font
\usepackage{helvet}         % selects Helvetica as sans-serif font
\usepackage{courier}        % selects Courier as typewriter font
\usepackage{type1cm}        % activate if the above 3 fonts are 
\usepackage{makeidx}         % allows index generation
\usepackage{graphicx}        % standard LaTeX graphics tool
\usepackage{multicol}        % used for the two-column index
\usepackage[bottom]{footmisc}% places footnotes at page bottom
\usepackage{multirow} %added by Shang
\usepackage{adjustbox} %added by Shang
\usepackage{mdframed} %added by Shang
\usepackage{comment} %added by Shang
\usepackage{orcidlink} %added by Shang
\makeindex             % used for the subject index
\begin{document}

\title*{Unlocking Excellence: The Impact of Voucher Incentives on Cybersecurity Education}
\titlerunning{Impact of Voucher Incentives on Cybersecurity Education}
\author{Jianhua Li, Shang Gao, Michelle Harvey, and Trina Myers}
\authorrunning{Li, et al.}
\institute{ Jianhua Li\orcidlink{0000-0001-5878-6032}, Shang Gao\orcidlink{0000-0002-2947-7780}, and Trina Myers\orcidlink{0000-0002-0252-0320} \at School of Information Technology, %, Deakin University, 75 Pigdons Rd, Waurn Ponds VIC 3216 %\email{\{jack.li,shang.gao,trina.myers\}@deakin.edu.au}
\and Michelle Harvey\orcidlink{0000-0002-4047-7845} \at School of Life and Environmental Sciences, \\\\Deakin University, 75 Pigdons Rd, Waurn Ponds, VIC 3216, Australia\\
\email{{jack.li, shang.gao, michelle.harvey, trina.myers}@deakin.edu.au}}
\maketitle

\abstract*{
While voucher incentives have been popular for primary and secondary schools, they are less used in higher education. In this study, we leverage industry voucher incentives to inspire students in cybersecurity education (CSE). We adopt a 100\% portfolio-based assessment strategy, where students can freely select their target grades in the investigated unit. We purposely design one of the high distinction (HD) tasks to be obtaining an industry certificate and provide vouchers to those who can accomplish a predefined set of tasks before a midpoint. The voucher recipients will use the voucher to access the industry certificate training materials and sit the certificate exam for free. Passing the certificate exam is one of the conditions for gaining an HD grade. Our survey and interviews reveal a substantial influence of voucher incentives on students' career aspirations. 
In light of the findings, recommendations on adopting voucher incentives in CSE or broader ICT education are offered for institutions and researchers. 
} 

\abstract{While voucher incentives have been popular for primary and secondary schools, they are less used in higher education. In this study, we leverage industry voucher incentives to inspire students in cybersecurity education (CSE). We adopt a 100\% portfolio-based assessment strategy, where students can freely select their target grades in the investigated unit. We purposely design one of the high distinction (HD) tasks to be obtaining an industry certificate and provide vouchers to those who can accomplish a predefined set of tasks before a midpoint. The voucher recipients will use the voucher to access the industry certificate training materials and sit the certificate exam for free. Passing the certificate exam is one of the conditions for gaining an HD grade. Our survey and interviews reveal a substantial influence of voucher incentives on students' career aspirations. 
In light of the findings, recommendations on adopting voucher incentives in CSE or broader ICT education are offered for institutions and researchers. 
}


\section{Introduction}
In the ever-evolving landscape, the demand for cybersecurity professionals has surged to unprecedented heights. According to the Australian Cyber Security Growth Network, there will be at least 17,000 additional cybersecurity workers needed by 2026.\footnote{https://www.austcyber.com/resources/sector-competitiveness-plan-2019/chapter3} As cyber threats continue to grow in complexity and frequency \cite{li2021too}, educational institutions face the crucial task of preparing a new generation of cybersecurity experts capable of defending against an array of digital adversaries. To meet this demand and equip students with the skills and knowledge necessary to safeguard digital environments, educational programs in cybersecurity have become increasingly vital \cite{aldaajeh2022role, newhouse2017national}.

Rooted in computer science, students may perceive cybersecurity as a complex and technical subject that requires advanced knowledge of computer systems and programming languages. The perceived difficulty of the subject matter can deter students from pursuing cybersecurity programs \cite{ajzen1986prediction}. If the enrolled students cannot relate curricula to career or academic goals, they may not dedicate themselves to their studies. Under these circumstances, motivating students and facilitating appropriate learning environments are critical for fostering their interests and learning outcomes. Although numerous motivation strategies for learning have been developed and elaborated in general programs \cite{schiefele1991interest}, motivating students to learn cybersecurity is less explored. 

In higher education, universities and colleges are exploring innovative strategies to motivate and engage students \cite{averill2020motivates,li2023current}. % 
One such strategy is the integration of industry certifications as an integral component of academic curricula \cite{shim2021integration}. These certifications, often administered by industry-recognized organizations, enable students to validate their skills and competencies in real-world scenarios \cite{paterson2022motivation}. While integrating industry certifications into educational programs is not a novel concept \cite{fenton2019integrating}, the methodological aspect of their impact on students' goal achievement remains less examined.

Industry voucher incentives refer to benefits provided to students in the form of vouchers. This incentive is designed to motivate students, recognize their achievements, and boost their career development. In this study, we employ industry voucher incentives as one of the motivation strategies in cybersecurity education (CSE) \cite{vsvabensky2020cybersecurity} and evaluate the methodological aspect of their impact on students' goal achievement. 

Our contributions are threefold. 
\begin{enumerate}
\item Uncover the multifaceted influence of the industry voucher incentives on students' academic aspirations, engagement, and overall performance within the context of portfolio-based assessments. Specifically, we seek to understand how the availability of voucher incentives influences students' academic motivation and career aspirations in cybersecurity education.
\item Contribute to a deeper understanding of how such incentives can enhance students' educational experiences and outcomes. Our research could provide valuable insights for educators and institutions seeking innovative approaches to motivate and engage students in CSE and broader ICT education. 
\item Clarify the potential impact of voucher incentives on developing students' essential skills and competencies demanded by the cybersecurity industry.
\end{enumerate}

\begin{figure}
    \centering
    \includegraphics[width=0.95\columnwidth]{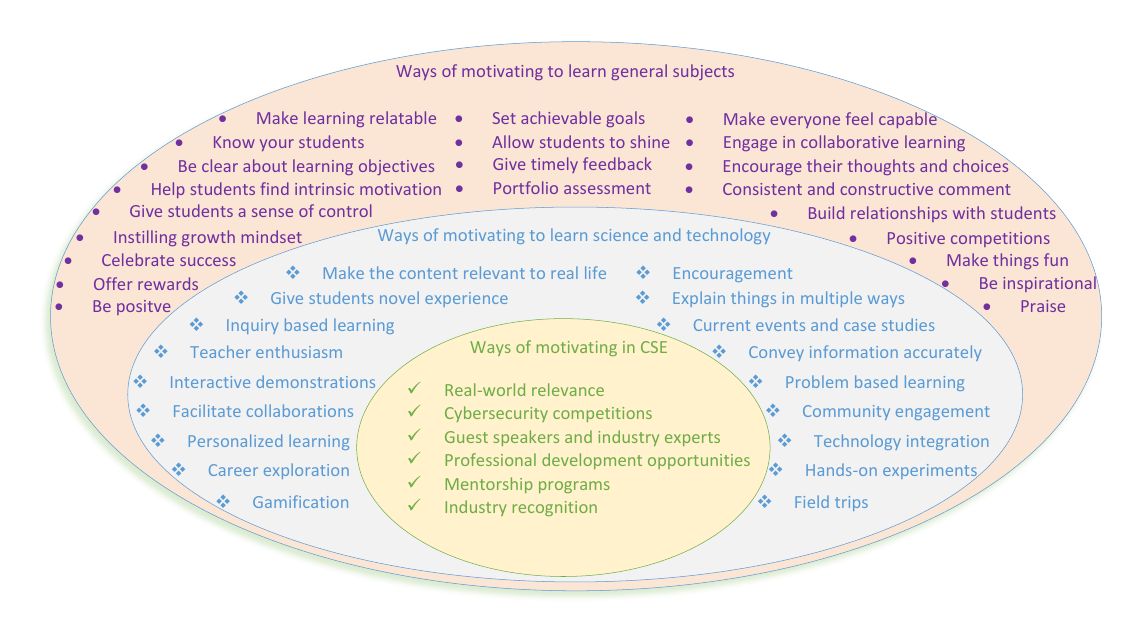}
    \caption{Motivation methods}
    \label{fig:motimeth}
\end{figure}

The remainder of this paper is organised as follows:  Section \ref{sec: literature review} presents the literature review on challenges in CSE and existing motivation methods, followed by an elaboration of our research questions and methodology in Section \ref{sec: research questions and methodology}. Section \ref{sec: findings} outlines the findings of employing voucher incentives in CSE. Section \ref{sec: discussion and recomendations} offers a comprehensive discussion of the results, our recommendations, and limitations. Finally, Section \ref{sec: conclusion} concludes this paper.

\section{Literature Review}
\label{sec: literature review}
\subsection{Challenges in CSE}
The growing importance of CSE in the digital age has been widely recognized in our work and daily life \cite{rahman2020importance, aldaajeh2022role}. Despite the high demand for cybersecurity professionals, CSE faces challenges in motivation. Rooted in computer science, students may consider cybersecurity as a complicated and technical subject requiring advanced mathematical and programming skills. Fear of not being able to master cybersecurity skills can prevent students from choosing this program \cite{giel2020fear}. On the other hand, some students may perceive a degree program as offering limited industry connections and outdated curricula \cite{raj2019cybersecurity}. Such negative perceptions can hinder students from completing their programs \cite{giannakos2017understanding}.

%importance of CSE and its challenges
\subsection{Motivation methods}
It is of paramount importance to motivate students to learn especially to those who show little interest \cite{schutte2015discover}. Previous studies have investigated a myriad of ways to motivate students to learn science and technology. 

For instance, Palmer comprehensively explored motivational strategies including success, novelty, choice, relevance, variety, collaboration, teacher enthusiasm, and encouragement \cite{palmer2007best}. The author discussed how to implement the above strategies in classrooms to seek the best way of motivating students in science. The conclusion is that all should be used as often as possible regularly. Butler leveraged the gamification technique to engage and motivate students in computer science, where 85\% of the students believed that they were strongly engaged \cite{butler2016gamification}. It was observed that converting complicated concepts into games could increase motivation.

In CSE, professors usually incorporate hands-on activities, labs, and simulations to enable students to apply theoretical concepts in practical settings. Vykopal \textit{et al.} argued that scalable hands-on training could improve learning experience motivation \cite{vykopal2021scalable}. OConnor \textit{et al.} designed a cybersecurity course for undergraduates in which students enjoyed attacking themselves \cite{oconnor2021teaching}. Theory and hands-on balanced courses motivate students with an authentic experience, while challenges are noticeable. Challenge-based activities may require students to have a large amount of knowledge and skills, which can be impractical for average students. Other significant challenges include maintaining standardised classroom lab settings and technical failures during rehearsals. 

%
% \begin{table}
% \caption{Please write your table caption here}
% \label{tab:1}       % Give a unique label
% %
% % Follow this input for your own table layout
% %
% \begin{tabular}{p{2cm}p{2.4cm}p{2cm}p{4.9cm}}
% \hline\noalign{\smallskip}
% Classes & Subclass & Length & Action Mechanism  \\
% \noalign{\smallskip}\svhline\noalign{\smallskip}
% Translation & mRNA$^a$  & 22 (19--25) & Translation repression, mRNA cleavage\\
% Translation & mRNA cleavage & 21 & mRNA cleavage\\
% Translation & mRNA  & 21--22 & mRNA cleavage\\
% Translation & mRNA  & 24--26 & Histone and DNA Modification\\
% \noalign{\smallskip}\hline\noalign{\smallskip}
% \end{tabular}
% $^a$ Table foot note (with superscript)
% \end{table}

\begin{table*}[ht]
\caption{Role of voucher incentives in education}
\label{tab:comp}
\begin{adjustbox}{width=\linewidth}
%\begin{tabular}{|l|l|l|}
\begin{tabular}{*3l}
\hline
\textbf{Aspect}                    & \textbf{General Voucher}                                                                                        & \textbf{Industry Voucher}                                                                                                          \\ \hline
\textbf{Focus}                     & Broad range of educational programs and services                                                       & Specific to industry-related certifications and training                                                                  \\ \hline
\textbf{Purpose}                   & Support diverse educational needs and goals                                                            & Enhance employability and skills in targeted industries                                                                   \\ \hline
\textbf{Eligibility}               & Available to students across various disciplines                                                       & \begin{tabular}[c]{@{}l@{}}Targeted towards students pursuing careers in specific \\ industries\end{tabular}              \\ \hline
\textbf{Coverage}                  & May include a variety of educational expenses                                                          & \begin{tabular}[c]{@{}l@{}}Typically covers costs related to certification exams, \\ training, and materials\end{tabular} \\ \hline
\textbf{Certifications Offered}    & Varied, may include academic, vocational, etc.                                                         & \begin{tabular}[c]{@{}l@{}}Industry-recognized certifications in specific fields \\ (e.g., IT, healthcare)\end{tabular}   \\ \hline
\textbf{Industry Alignment}        & Not necessarily aligned with specific industries                                                       & Aligned with industry standards and requirements                                                                          \\ \hline
\textbf{Cost Reduction}            & Helps reduce financial barriers to education                                                           & \begin{tabular}[c]{@{}l@{}}Reduces the financial burden of obtaining industry \\ certifications\end{tabular}              \\ \hline
\textbf{Employability Enhancement} & Enhances general employability across sectors                                                          & \begin{tabular}[c]{@{}l@{}}Increases job prospects and advancement opportunities \\ in targeted industries\end{tabular}   \\ \hline
\textbf{Skill Enhancement}         & Provides general knowledge and skills development                                                      & \begin{tabular}[c]{@{}l@{}}Offers specialized training and expertise in industry\\ -specific areas\end{tabular}           \\ \hline
\textbf{Professional Networking}   & May offer networking opportunities across sectors                                                      & \begin{tabular}[c]{@{}l@{}}Provides access to industry-specific networks and \\ communities\end{tabular}                  \\ \hline
\textbf{Career Advancement}        & Enhances overall career prospects and opportunities                                                    & \begin{tabular}[c]{@{}l@{}}Facilitates advancement in targeted industries and \\ professions\end{tabular}                 \\ \hline
\textbf{Employer Recognition}      & Offers recognition for academic achievements                                                           & \begin{tabular}[c]{@{}l@{}}Demonstrates proficiency and expertise valued by \\ employers\end{tabular}                     \\ \hline
\textbf{Lifelong Learning Support} & \begin{tabular}[c]{@{}l@{}}Encourages continuous learning and professional \\ development\end{tabular} & \begin{tabular}[c]{@{}l@{}}Supports ongoing skill enhancement and certification \\ renewal\end{tabular}                   \\ \hline
\textbf{Impact on Industry Needs}  & \begin{tabular}[c]{@{}l@{}}May address general workforce demands and \\ shortages\end{tabular}         & \begin{tabular}[c]{@{}l@{}}Addresses specific skill gaps and demands in targeted \\ industries\end{tabular}               \\ \hline
\end{tabular}
\end{adjustbox}
\end{table*}

Competition-based learning demonstrates its effectiveness in motivating students in terms of attracting students and enhancing learning outcomes. Khan \textit{et al.} investigated the role of competition in delivering CSE at the United Arab Emirates University \cite{khan2022game}. They designed a course following the attention, relevance, confidence and satisfaction (ARCS) model. In a broader context, Li \textit{et al.} reviewed the use of the ARCS model in computer-assisted learning across multiple countries and school levels \cite{li2018use}. The authors confirmed that the model played a positive role in motivating learners in diverse cases. However, Cheung \textit{et al.} researched challenge-based learning in CSE and found that the attendance to cybersecurity classes depended largely on individual interest in the subject \cite{cheung2011challenge, cheung2012effectiveness}. Fig. \ref{fig:motimeth} presents the commonly used motivation ways, starting from general subjects, and narrowing down to science and technology and CSE subjects, as identified in our study.

\subsection{Research gaps}
Although motivated students tend to have better academic performance, a particular motivation method may be effective for some students but ineffective for others. Several factors can impact an individual's motivation level, including personal goals, expectancy, values and beliefs. Previous works for CSE inadequately consider such individual attributes. To this end, a portfolio-based assessment (PBA) strategy could provide personalized learning to meet various demands \cite{yastibas2015use}. 

This PBA strategy often involves ongoing feedback and reflection, rather than relying solely on summative assessments. This continuous loop provides more opportunities for professors to engage and motivate students progressively. More importantly, this personalized approach to assessments can enhance students' sense of ownership and autonomy over their learning process, leading to increased motivation. However, there is a lack of publications showcasing the PBA's role in motivating cybersecurity undergraduates.

Vouchers have been widely used in education for many years, where educational vouchers are typically issued by governments to reinforce freedom of choice, productive efficiency, equity and social cohesion \cite{levin2002comprehensive, west1996education, jongbloed2000vouchers}. Unlike traditional educational vouchers as an alternative for financing purposes, industry voucher incentives emerge as a promising strategy to enhance career development opportunities for students. 

In detail, industry certification bodies or educational institutions, offer students access to training materials, certification exams, and other resources at reduced or no cost. By incentivizing students to pursue industry-recognized certifications, vouchers are used to bridge the gap between academic learning and industry expectations, validate students' skills and knowledge, and increase their employability. Table \ref{tab:comp} presents our comparative findings in this study. 

As little research has been done to investigate vouchers' role in motivating students \cite{jongbloed2000vouchers}, this study aims to address these gaps by investigating the role of voucher incentives in boosting career aspirations for students in CSE.

\section{Research Questions and Methodology}
\label{sec: research questions and methodology}
Based on the above discussions, we identify the following three primary questions central to this study:
\begin{enumerate}
\item Does the voucher incentive motivate students to pursue academic excellence in CSE, and/or influence individual study habits or levels of engagement?
\item Does the voucher incentive influence career aspirations and/or enhance industry recognition?
\item What additional support or resources could educational institutions provide to enhance the voucher incentive effectiveness?
\end{enumerate}

\subsection{Study Method}
We select a year-one entry-level cybersecurity major unit and develop a 100\% portfolio-based assessment model to replace traditional evaluation forms (i.e., assignments, tests). This portfolio-based model offers a more holistic, authentic, and personalized approach to evaluating student learning, promoting deeper engagement, and preparing students for the challenges of theoretical knowledge and practical cybersecurity skills. 

In this investigated unit, students can select their target grades from Pass to High Distinction (HD) at the beginning of a trimester. The higher the target selected, the more challenging tasks to complete. For example, if HD, the highest grade, is selected, the HD seekers must have completed all the tasks designed for this unit to secure their HD grade. We purposely integrate a widely-recognized industry certification as one of the HD tasks. To complete this HD task, HD seekers need to first perform well on a set of voucher-eligible tasks, then independently study all the certificate-relevant training materials, and finally pass the certificate exam provided by the certification body within a given period (e.g., 6 weeks).

\begin{figure}
    \centering
    \includegraphics[width=0.7\columnwidth]{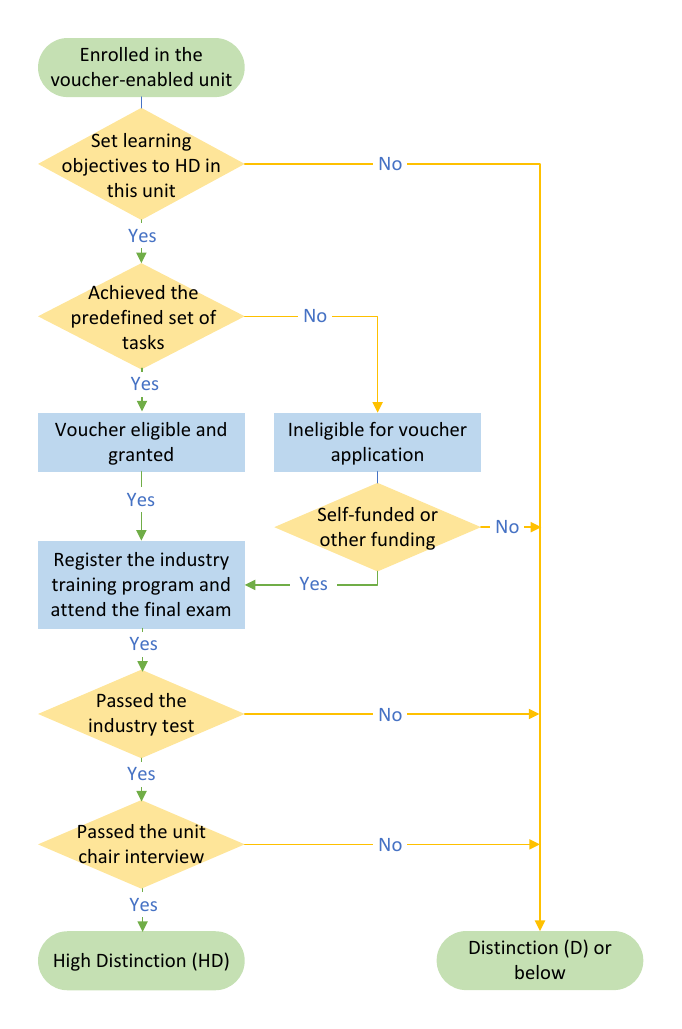}
    \caption{Voucher granting and HD achieving process}
    \label{fig:voucherHD}
\end{figure}

Central to this educational model is leveraging voucher incentives on students’ academic aspirations within the Bachelor of Cybersecurity. Designed to challenge and reward high-achieving students who aspire to excel in their cybersecurity studies, the voucher incentive aligns with pursuing HD within the academic portfolio. The industry certification exam and training materials are free for students with the granted voucher. This voucher is accessible to students who achieve all the predefined set of tasks before a midpoint deadline, providing additional motivation and recognition for their dedication to academic excellence. The predefined voucher-eligible task set includes 11 assignments (1 HD, 1 D, 2 C and 7 P tasks) while its deadline is set to the start of the middle week of each trimester (e.g., the 6th week out of 11 weeks).

The voucher information is readily available on the educational institution’s official website for openness, transparency and equity. Initially, we teaching staff communicated to our students that the voucher program would serve as a reward designed to motivate exceptional academic achievements and boost career aspirations. Thanks to the portfolio-based assessment approach, students can select their target grades based on individual needs. 

Fig. \ref{fig:voucherHD} depicts the voucher granting and HD achieving process. Students need to set their target to HD and attempt the predefined voucher-eligible tasks before the set date. If the attempt quality meets the requirements, the students receive the voucher. Otherwise, they can fund themselves or secure other funding sources for this industry training program. Once passed the certificate exam before the due date of the portfolio which is usually set to the end of the trimester, the students will be interviewed and the HD grade can be granted when eligible. Otherwise, the student's grades will fall back to D or lower.

%\subsection{Study Design}

\subsection{Data Collection}
In this study, we employ mixed approaches to comprehensively investigate the impact of voucher incentives on student academic achievement in CSE. Both quantitative and qualitative data collection and analysis techniques are used to scrutinize the research phenomenon.

\textbf{Surveys.}
To capture quantitative data, a survey instrument was designed and administered to students who had completed the unit. We built a survey site on the Qualtrics platform and provided the survey link on the unit sites to all 435 students in 2022 and 2023, including those who had received vouchers and those who had not. As this survey was voluntary and anonymous, 24 participants completed this online survey. Although the sample size was relatively small, we successfully captured data from a variety of students, including those who elected to study this unit and those for whom it was mandatory. Specifically, this unit is a core unit for a Bachelor of Cybersecurity (taken in Year 1) and an elective for other undergraduates (taken in Year 2 or Year 3). The collected information focused on academic goal-setting, career aspiration, and the perceived impact of vouchers on academic achievements.

\textbf{Qualitative Interviews.}
Qualitative follow-up interviews were conducted with a subset of students. While ten students were randomly invited via email, we were able to get five volunteer interviewees. Email interviews were adopted to afford the participants ample time for reflection and to formulate well-considered responses. These interviews provided an in-depth exploration of their experiences with voucher incentives and their perceptions of how vouchers influenced their academic and career trajectories. Open-ended questions were posed to elicit rich and detailed responses, allowing for deeper insights into the qualitative aspects of the research.

\subsection{Data Analysis}

%\subsubsection{Quantitative Analysis}
Quantitative data collected through surveys were subjected to rigorous statistical analysis \cite{romero2010educational}. Descriptive statistics, like percentages, were calculated to summarize survey responses. Comparative analysis, such as cross-tabulations, was employed to identify patterns and trends among students across multiple campuses.

%\subsubsection{Qualitative Analysis}
Qualitative data from interviews were analyzed using thematic analysis \cite{lemay2021comparison}. Collected interview responses were systematically coded to identify recurring themes and patterns within the data. Themes related to academic goal-setting, motivation, study habits, and career aspirations were extracted to gain a deeper understanding of the qualitative dimensions of this research.

\subsection{Ethical Considerations}

Ethical considerations are of paramount importance throughout the research process. Participation in the survey and interview was entirely voluntary, and informed consent was obtained from all participants. Additionally, participants were assured of the anonymity and confidentiality of their responses. Ethical approval for this research was obtained from the university's ethics review board with a reference code SEBE-2023-48.

% \subsection{Limitations}

% While this mixed-methods case study approach offers a comprehensive perspective on the research phenomenon, it is essential to acknowledge certain limitations. The study's findings may be context-specific to the university and the investigated unit, limiting generalizability. Furthermore, the research design is retrospective, capturing students' perceptions after completing their units, which may introduce recall bias \cite{blome2015measuring}.

\section{Findings}
\label{sec: findings}
\subsection{Descriptive Statistics of Survey Data}
\subsubsection{Participants Distribution}
\begin{figure}[ht!]
\sidecaption[t]
\includegraphics[scale=.7]{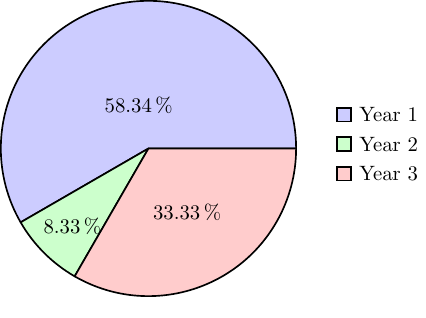}
%
% If no graphics program available, insert a blank space i.e. use
%\picplace{5cm}{2cm} % Give the correct figure height and width in cm
\caption{Distribution of participants. The investigated unit is a foundational  unit for Year 1 students enrolled in the Bachelor of Cybersecurity program, while it serves as an elective for other students. This distribution underscores the diversity of our data samples.}
\label{fig:year}       % Give a unique label
\end{figure}

As previously explained, the investigated unit is typically taken by first-year students pursuing a Bachelor of CyberSecurity as a core, although it's open to second and third-year students in other programs as an elective. Notably, such a degree program spans three years at the university. Fig. \ref{fig:year} illustrates the distribution of participants in this study, with 58.34\% from Year 1, 8.34\% from Year 2, and 33.33\% from Year 3. Two-thirds of participants said they received at least one voucher in their program while 90.91\% of them gained the investigated voucher. This diverse mix of participants facilitated a comprehensive exploration of the voucher's impact on various aspects, namely, as a compulsory or an elective unit.

% \begin{figure}[ht!]
%     \centering
%     % Include the figure
%     \includegraphics[width=0.2\textwidth]{1year_distribution.pdf}\\
%     The investigated unit is a foundational  unit for Year 1 students enrolled in the Bachelor of Cybersecurity program, while it serves as an elective for other students. This distribution underscores the diversity of our data samples.
%     % Add a caption
%     \caption{Distribution of participants}
%     % Add a label for referencing
%     \label{fig:year}
%     % Add a comment about the unit
    
% \end{figure}

\subsubsection{Goal-Setting and Academic Aspirations}
\begin{figure}[h]
%\sidecaption[h]
\centering
\includegraphics[scale=.7]{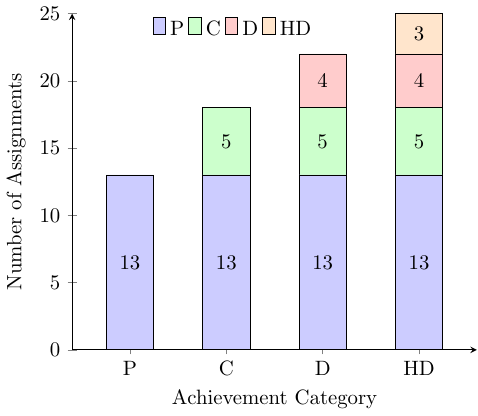}
%
% If no graphics program available, insert a blank space i.e. use
%\picplace{5cm}{2cm} % Give the correct figure height and width in cm
\caption{Distribution of various leveled assignments}
\label{fig:task}     % Give a unique label
\end{figure}

In our study, students have the opportunity to set academic target grades, ranging from pass (P), credit (C), and distinction (D) to high distinction (HD). These goals entail completing different levels of assignments within their academic portfolios. Fig. \ref{fig:task} outlines the task requirements for each grade. To clarify, a student must complete 13 P-level tasks to pass the unit, while an HD aspirant needs to accomplish up to 25 tasks within 11 weeks.

% \begin{figure}
%     \centering
%     \includegraphics{2task_distribution.pdf}
%     \caption{Distribution of various levels of assignments}
%     \label{fig:task}
% \end{figure}

% \begin{figure}[h]
%     \centering
%     \includegraphics[width=0.2\textwidth]{2task_distribution.pdf}
%     \caption{Distribution of various leveled assignments}
%     \label{fig:task}
% \end{figure}

\begin{table}[ht]
\caption{Distribution of students with  target grades in 2023}
%\begin{adjustbox}{width=0.6\linewidth}
%\begin{tabular}{|l|l|l|l|l|l|}
\begin{tabular}{*6l}
\hline
& \begin{tabular}[c]{@{}l@{}}\textbf{No. of} \\ \textbf{students}\end{tabular}       & \begin{tabular}[c]{@{}l@{}}\textbf{Pass}\\ \textbf{seekers}\end{tabular} & \begin{tabular}[c]{@{}l@{}}\textbf{Credit}\\ \textbf{seekers}\end{tabular} & \begin{tabular}[c]{@{}l@{}}\textbf{Distinction}\\ \textbf{seekers}\end{tabular} & \begin{tabular}[c]{@{}l@{}}\textbf{HD}\\ \textbf{seekers}\end{tabular} \\ \hline
\multirow{2}{*}{\begin{tabular}[c]{@{}l@{}}\textbf{Urban}\\ \textbf{campus}\end{tabular}}    & 182                                                              & 70                                                     & 50                                                       & 30                                                            & 32                                                   \\ \cline{2-6} 
& \begin{tabular}[c]{@{}l@{}}Pecentage\\ distribution\end{tabular} & 38.47\%                                                & 27.47\%                                                  & 16.48\%                                                       & 17.58\%                                              \\ \hline
\multirow{2}{*}{\begin{tabular}[c]{@{}l@{}}\textbf{Cloud}\\ \textbf{campus}\end{tabular}}    & 55                                                               & 23                                                     & 13                                                       & 8                                                             & 11                                                   \\ \cline{2-6} 
& \begin{tabular}[c]{@{}l@{}}Pecentage\\ distribution\end{tabular} & 41.82\%                                                & 23.64\%                                                  & 14.54\%                                                       & 20.00\%                                              \\ \hline
\multirow{2}{*}{\begin{tabular}[c]{@{}l@{}}\textbf{Regional}\\ \textbf{campus}\end{tabular}} & 35                                                               & 19                                                     & 11                                                       & 3                                                             & 1                                                    \\ \cline{2-6} 
& \begin{tabular}[c]{@{}l@{}}Pecentage\\ distribution\end{tabular} & 54.29\%                                                & 31.43\%                                                  & 8.57\%                                                        & 2.86\%                                               \\ \hline
\multirow{2}{*}{\textbf{Total}}                                                     & 272                                                              & 112                                                    & 74                                                       & 41                                                            & 44                                                   \\ \cline{2-6}  & \begin{tabular}[c]{@{}l@{}}Pecentage\\ distribution\end{tabular} & 41.18\%                                                & 27.21\%                                                  & 15.07\%                                                       & 16.18\%                                              \\ \hline
\end{tabular}
%\end{adjustbox}
\label{tab:goal}
\end{table}

Table \ref{tab:goal} presents a comprehensive overview of students' target grades in 2023. Notably, within the Urban campus, 38.47\%  of students aimed to attain a ``P,'' 27.47\% targeted ``C,'' 16.48\% aspired to ``D,'' and 17.58\% set their sights on achieving ``HD.'' In contrast to students from Cloud or Urban campuses \cite{clift2016educational}, Regional students \cite{fernandez2017parental} demonstrated a different pattern, with over 54\% striving for a ``Pass'' grade, while approximately 11\% pursued either ``D'' or ``HD''. Likewise, Table \ref{tab:goal1} demonstrates the distribution in 2022, which displays a similar pattern.

\begin{table}[ht]
\caption{Distribution of students with  target grades in 2022}
%\begin{adjustbox}{width=0.6\linewidth}
%\begin{tabular}{|l|l|l|l|l|l|}
\begin{tabular}{*6l}
\hline
%& \begin{tabular}[c]{@{}l@{}}No. of \\ students\end{tabular}       & \begin{tabular}[c]{@{}l@{}}Pass\\ seekers\end{tabular} & \begin{tabular}[c]{@{}l@{}}Credit\\ seekers\end{tabular} & \begin{tabular}[c]{@{}l@{}}Distinction\\ seekers\end{tabular} & \begin{tabular}[c]{@{}l@{}}HD\\ seekers\end{tabular} \\ \hline
& \begin{tabular}[c]{@{}l@{}}\textbf{No. of} \\ \textbf{students}\end{tabular}       & \begin{tabular}[c]{@{}l@{}}\textbf{Pass}\\ \textbf{seekers}\end{tabular} & \begin{tabular}[c]{@{}l@{}}\textbf{Credit}\\ \textbf{seekers}\end{tabular} & \begin{tabular}[c]{@{}l@{}}\textbf{Distinction}\\ \textbf{seekers}\end{tabular} & \begin{tabular}[c]{@{}l@{}}\textbf{HD}\\ \textbf{seekers}\end{tabular} \\ \hline
\multirow{2}{*}{\begin{tabular}[c]{@{}l@{}}\textbf{Urban}\\ \textbf{campus}\end{tabular}}    & 75                                                              & 43                                                     & 13                                                       & 6                                                            & 13                                                   \\ \cline{2-6} 
& \begin{tabular}[c]{@{}l@{}}Pecentage\\ distribution\end{tabular} & 57.33\%                                                & 17.33\%                                                  & 8.00\%                                                       & 17.33\%                                              \\ \hline
\multirow{2}{*}{\begin{tabular}[c]{@{}l@{}}\textbf{Cloud}\\ \textbf{campus}\end{tabular}}    & 75                                                               & 37                                                     & 15                                                       & 7                                                             & 16                                                   \\ \cline{2-6} 
& \begin{tabular}[c]{@{}l@{}}Pecentage\\ distribution\end{tabular} & 49.33\%                                                & 20.00\%                                                  & 9.33\%                                                       & 21.33\%                                              \\ \hline
\multirow{2}{*}{\begin{tabular}[c]{@{}l@{}}\textbf{Regional}\\ \textbf{campus}\end{tabular}} & 13                                                               & 9                                                     & 2                                                       & 1                                                             & 1                                                    \\ \cline{2-6} 
& \begin{tabular}[c]{@{}l@{}}Pecentage\\ distribution\end{tabular} & 69.23\%                                                & 15.38\%                                                  & 7.69\%                                                        & 7.69\%                                               \\ \hline
\multirow{2}{*}{\textbf{Total}}                                                     & 163                                                              & 89                                                    & 30                                                       & 14                                                            & 30                                                   \\ \cline{2-6}  & \begin{tabular}[c]{@{}l@{}}Pecentage\\ distribution\end{tabular} & 54.60\%                                                & 18.40\%                                                  & 8.59\%                                                       & 18.40\%                                              \\ \hline
\end{tabular}
%\end{adjustbox}
\label{tab:goal1}
\end{table}

The variance in academic aspirations among Regional students may be attributed to unique circumstances or factors that influence their goals \cite{bonnor2021structural}, such as limited access to resources and alignment with local job markets. % \cite{li2023adapting}  
Further research may be warranted to delve deeper into the specific challenges and potential solutions that Regional students encounter. However, an extensive exploration of this topic is beyond the scope of the present paper.

The survey data uncovered noteworthy trends related to academic goal-setting. Students striving for higher distinctions tended to engage more actively with the unit's content and assignments. The clearly defined academic goal-setting framework played an important role in enhancing students' sense of purpose and motivation.

\subsubsection{Granted Voucher}
Despite the absence of limitations on the number of funded vouchers, securing one remains a challenging endeavour. In 2023, 18 out of 182 Urban students, 12 out of 55 Cloud students, and 1 out of 35 Regional students were awarded vouchers. This equates to a success rate of 9.89\%, 21.82\%, and 2.86\% for Urban, Cloud, and Regional students, respectively. On average, 44 out of 435 students, or approximately 10.11\% of all students in 2022 and 2023, successfully earned the voucher.

\subsubsection{Impact on Career Developments (Voucher Recipients)}
Subsequently, in terms of the impact on career development among voucher recipients, our survey data revealed the following:
\begin{itemize}
    \item A vast majority of voucher recipients (85.33\%) were highly motivated to aim for an HD in the associated unit(s) due to the availability of vouchers.
    \item Voucher recipients consistently reported positive effects on their career aspirations, with 66.67\% acknowledging such influence.
    \item When asked to rate the overall impact of the vouchers on their academic achievements, a significant portion (75\%) characterized it as either ``Very positive'' or ``Positive.''
\end{itemize}

%\begin{comment}
\subsection{Insights Gained from Interviews}
%A follow-up interview was conducted with volunteers, focusing on motivation, student engagement, academic achievement, industry recognition, support, and overall experience. 
We summarize our findings and reflect on the interviewees' comments. Please note that their transcripts are included in blue-lined frames for better readability.
\subsubsection{Motivation and Academic Goals}

Regarding their motivations for pursuing HD, one interviewee shared, \\

\begin{mdframed}[linecolor=blue]
\textit{``The voucher was the single biggest motivator for aiming to achieve a high distinction grade. As someone who is about to enter the workforce, I know how valuable industry certifications are, and being able to complete this unit with a voucher to sit an industry certification exam is very valuable.''}
\end{mdframed}

\vspace{0.2cm}

\textbf{Reflection 1a:} This comment underscores the powerful incentive that vouchers represent for students, particularly those on the cusp of transitioning to their professional careers. It highlights the recognition of the practical value of industry certifications and how they align with students' career aspirations.

Another student commented, \\
\begin{mdframed}[linecolor=blue]
\textit{``Because I wanted the voucher for the certification, and only students aiming for a high distinction were eligible to receive a voucher, I aimed for a high distinction.''}
\end{mdframed}

\vspace{0.2cm}

\textbf{Reflection 1b:} This statement illustrates the strategic role that voucher eligibility criteria play in shaping academic goals. It showcases how the opportunity to access vouchers motivates students, as well as guides their goal-setting within the academic context.

\subsubsection{Changes in Study Habits and Engagement}
We asked interviewees to share examples of how their study habits changed because of the voucher incentive. Some interviewees said, \\

\begin{mdframed}[linecolor=blue]
\textit{``I had to increase my study habits to ensure that I was completing all the tasks that I needed to complete to receive the voucher.''}
\end{mdframed}

\vspace{0.2cm}

\textbf{Reflection 2a:} These responses highlight a significant shift in study habits driven by the voucher incentive. The need to complete specific tasks for voucher eligibility acted as a catalyst for increased dedication to coursework. It demonstrates how external motivators can influence students to adopt more disciplined and focused study routines.

Another student mentioned, \\
\begin{mdframed}[linecolor=blue]
\textit{``I had to complete more of the unit’s tasks, which therefore naturally led to increased engagement with the unit. Without the enticement of the voucher, my engagement with the unit may have been less.''}
\end{mdframed}

\vspace{0.2cm}

\textbf{Reflection 2b:} This comment underscores the symbiotic relationship between voucher incentives and student engagement. It suggests that the prospect of earning a voucher led to greater involvement in the unit's tasks, enhancing the overall learning experience. It highlights how incentives can enhance academic performance, as well as the quality of engagement with the subject matter.

\subsubsection{Academic Achievement and Self-Perception:}
Vouchers motivated students to aim high in their studies. An interviewee admitted, \\

\begin{mdframed}[linecolor=blue]\textit{``I completed all of the high distinction tasks to a high level so that I would get them marked as completed.''}
\end{mdframed}

\vspace{0.2cm}

\textbf{Reflection 3:} This interviewee's statement reveals the extent to which voucher incentives can inspire students to elevate their academic ambitions. The pursuit of HD tasks to a high standard illustrates a proactive approach to learning, driven by the desire to secure the voucher. This commitment not only impacts academic achievement but also shapes students' self-perception as motivated and goal-oriented learners.

\subsubsection{Career Aspirations and Industry Recognition}
Thanks to the industrial endorsement, students are enthusiastically pursuing vouchers to advance their careers in cybersecurity. For instance, an interviewee expressed,\\
\begin{mdframed}[linecolor=blue]\textit{``Yes, as someone wanting to enter into the cybersecurity industry within the next 12 months, the fact that I have been offered this voucher has been excellent and helps to make me more competitive in the workforce.''}
\end{mdframed}

\vspace{0.2cm}

\textbf{Reflection 4a:} This interviewee's response underscores the direct link between the voucher incentive and students' career aspirations. It highlights how the opportunity to obtain industry-recognized certifications through vouchers enhances their competitiveness in the job market, aligning their academic pursuits with their professional goals.

Others commented, \\
\begin{mdframed}[linecolor=blue]
\textit{``It is a very well-known certification and it will help me to get up-to-date knowledge and hands-on practice. It is also helpful getting experience.''} 
\end{mdframed}
\vspace{0.2cm}
\begin{mdframed}[linecolor=blue]
\textit{``It shows my professionalism to the employers.''}
\end{mdframed}
\vspace{0.2cm}
\begin{mdframed}[linecolor=blue]
\textit{``I believe that having a certification is extremely valuable, as it will help differentiate high achieving students from their peers in the workforce.''}
\end{mdframed}

\vspace{0.2cm}

\textbf{Reflection 4b:} These comments collectively emphasize the multifaceted benefits of industry-recognized certifications obtained through vouchers. They suggest that such certifications not only provide knowledge and practical skills but also serve as a symbol of professionalism and competence. Additionally, they contribute to students' confidence in their future career prospects.

\subsubsection{Support and Resources}
Given the requirement for students to complete 25 tasks within 11 weeks, with a substantial portion of these tasks due within the initial 5 weeks to secure a voucher, the pursuit of HD can be exceptionally tough. In response to inquiries about the need for additional support and resources, an interviewee candidly remarked,\\
\begin{mdframed}[linecolor=blue]
\textit{``The only feedback which I could think of is to issue the vouchers closer to the end of week 5. Ideally, as close as possible.''} 
\end{mdframed}

\vspace{0.2cm}

\textbf{Reflection 5a:} It is evident that some students expressed a desire for the voucher allocation to occur at an earlier stage, providing them with more time for thorough preparation. This interviewee's suggestion reflects the practical challenges students may face when striving for HD in a compressed timeframe. Their input underscores the importance of carefully timed voucher issuance, which aligns with the critical task deadlines. It also points to the significance of responsiveness to student feedback in optimizing the voucher incentive program. Notably, 30\% of voucher recipients were willing to forgo the pursuit of an HD as they prioritized adequate preparation time for their industrial certification exams. This decision was driven by the simultaneous demands of multiple assignments in this unit, making it too risky to proceed with the exam without enough preparation.

In contrast, several interviewees voiced their concerns about the challenges of juggling concurrent tasks across multiple units, particularly during peak assignment periods. Given the demanding nature of meeting the voucher criteria, certain students articulated their need for enhanced support in terms of scheduling and preparing for certification exams. One student candidly stated, \\
\begin{mdframed}[linecolor=blue]
\textit{``The worst moment of my year was that I had 8 assignments due at the same time.''} 
\end{mdframed}

\vspace{0.2cm}

\textbf{Reflection 5b:} This student's comment sheds light on the challenging nature of academic life, where students often face the daunting task of managing multiple assignments simultaneously. It underscores the importance of considering the broader context in which voucher incentives operate, acknowledging the potential strain on students who are striving for excellence in several areas of study.

\subsubsection{Reflecting on the Experience}
Participants in the study were open about sharing their voucher experiences with prospective students, with one individual offering enthusiastic advice:\\ 
\begin{mdframed}[linecolor=blue]
\textit{``Do it! Your future self will thank you.''}
\end{mdframed}

\vspace{0.2cm}

\textbf{Reflection 6a:} It seems that this student is well-motivated and confident in his/her academic and career development.

Another student suggested,\\
\begin{mdframed}[linecolor=blue]
\textit{``To further enhance the industrial certification voucher program, our university can expand certification options, offer flexible exam scheduling, and create a centralized online resource hub. Skill assessments can help students choose the most suitable certifications, while industry partnerships can provide hands-on experience.''}
\end{mdframed}

\vspace{0.2cm}

\textbf{Reflection 6b:} These comments reflect students' active engagement with the voucher incentive program and their desire to see it continually evolve to better serve their needs. The advice given to prospective students underscores the program's positive impact on academic and career aspirations.

\begin{figure}[h]
%\sidecaption[t]
\centering
\includegraphics[scale=.7]{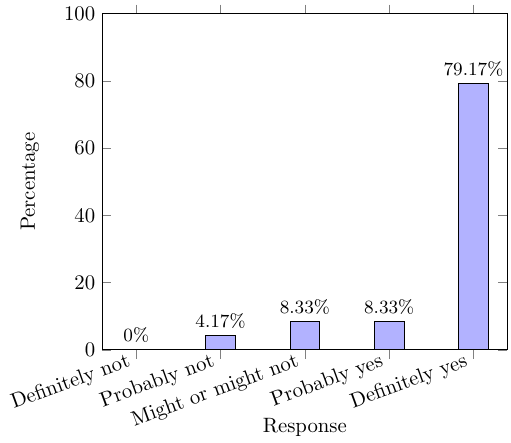}
\caption{Do you believe that universities should allocate more budget to support such vouchers?}
\label{fig:budget}      
\end{figure}

When the issue of potential budget limitations was raised, many students expressed their desire for increased financial support for voucher programs. Fig. \ref{fig:budget} provides a visual representation of this sentiment across all participants, with over 90\% of students indicating their hope for greater financial backing in this regard. One student further emphasized this point, stating:\\
\begin{mdframed}[linecolor=blue]
\textit{``Our university is one of the few Universities in Kings Field where they advertise that they offer industry certifications on their website. Other Uni's do not advertise/offer these certifications and our university should continue to differentiate itself in this manner.''}
\end{mdframed}

\vspace{0.2cm}

\textbf{Reflection 6c:} These reflections collectively highlight the program's success in motivating students and fostering their career ambitions. They also underscore the importance of addressing financial considerations to ensure the program's sustainability and continued positive impact.

% \begin{figure}[h]
%     \centering
%     \includegraphics[width=0.25\textwidth]{figures/7more_budget_hist.pdf}
%     \caption{Do you believe that universities should allocate more budget to support such vouchers?}
%     \label{fig:budget}
% \end{figure}

%\end{comment}

\section{Discussion and Limitation}
\label{sec: discussion and recomendations}
%\subsection{Industry Exam Scheduling}
%In Australian universities, a full-time postgraduate usually needs to complete four units within 12 weeks. An introduction of additional training materials and industry exams can be extremely challenging to students. early
%Regarding exam scheduling, our observations include the following:
%\begin{itemize}
%    \item Approximately 52\% of participants successfully acquired a voucher, 22\% did not, and 26\% chose not to disclose their access status.
%    \item In terms of exam venue preferences, 30\% of participants expressed a preference for the university to organize their exams, while 40\% indicated their intention to book the exam independently.
%    \item Notably, 30\% of voucher recipients were willing to forgo the pursuit of an HD as they prioritized adequate preparation time for their industrial certification exams. This decision was driven by the simultaneous demands of multiple assignments, making it too risky to proceed with the exam without enough preparation.
%\end{itemize}
%At the heart of this research lies the utilization of voucher incentives, with a particular focus on the Certified Secure Computer User (CSCU) voucher, as a reward for high-achieving students. The findings pertaining to voucher utilization illuminate its effectiveness.

%revised up here....

\subsection{Impact of Voucher Incentives on Academic Aspirations}
The findings of this study highlight the significant impact of industry voucher incentives on academic aspirations for the investigated unit. Voucher recipients were highly motivated to aim for the highest target, an HD, due to the availability of vouchers. This motivation is a promising indicator of the effectiveness of voucher incentives in fostering excellence in academic achievement. The unit chairs received numerous requests from students who were eager to acquire the voucher, even if they did not initially meet the eligibility requirements. Remarkably, around ten students indicated their willingness to personally finance their industry certification exams to achieve their ``HD'' academic goal.

\textbf{Recommendation 1:} Given the positive influence of voucher incentives on academic aspirations, institutions may consider expanding the use of similar incentives within cybersecurity and related educational programs, encouraging students to seek higher academic goals for better learning outcomes \cite{belfield2015privatizing}.

\subsection{Enhancement of Student Engagement}
The survey data revealed that voucher recipients reported increased engagement and improved study habits. This finding suggests that voucher programs motivate students to excel and foster deeper involvement with course content and assignments.

\textbf{Recommendation 2:} Educators may actively promote the integration of voucher incentives as a methodological tool to enhance student engagement in cybersecurity and ICT education \cite{blais2021self}.

\subsection{Perceptions of Voucher Effectiveness}
Students’ irresistibly positive perceptions of voucher incentives as effective motivators underscore the potential of this methodological approach in educational research. Vouchers were tools that improved academic performance and motivation among students.

\textbf{Recommendation 3:} Institutions may consider conducting further research to scrutinize the long-term impact of voucher incentives on students' career trajectories and the extent to which these incentives contribute to improved employability within the cybersecurity industry.

\subsection{Role of Methodological Innovation}
This study demonstrates the transformative potential of voucher incentives as methodological tools in educational research. By integrating these incentives into the educational environment, educational researchers gain insights into the multifaceted impact on academic and career outcomes.

\textbf{Recommendation 4:} Educational researchers may continue to explore innovative methodological approaches, such as voucher incentives, to enhance the depth and breadth of their research in the field of CSE and beyond \cite{wickert2021management}.

\subsection{Generalizability and Further Research}
It is essential to acknowledge the context-specific nature of this study, conducted within the investigated unit at our university. Further research is needed to assess the generalizability of these findings in different educational contexts and disciplines.

\textbf{Recommendation 5:} Future research may consider conducting similar studies across diverse educational settings to evaluate the transferability of voucher incentive programs and their impact on student's academic and career aspirations \cite{neto2021automatic}.

This includes further investigating the application of methodological tools like voucher incentives in various educational contexts. Additionally, investigating the long-term impact of these methodological approaches on students' academic and career trajectories can provide further insights into their effectiveness.

\subsection{Recommendations for Schools/Institutions}
The following recommendations provide key points for schools and institutions to enhance the effectiveness and inclusivity of voucher incentive programs in CSE. By addressing issues in equity, curriculum relevance, training, and resource allocation, schools and institutions can create a supportive and equitable learning environment that empowers students to excel in their academic pursuits.
\begin{itemize}
    \item \textbf{Equity Considerations}: To ensure equitable access to voucher incentive programs, institutions should actively monitor their impact across diverse student populations. This monitoring should encompass factors such as gender, socioeconomic background, and geographical location \cite{bonnor2021structural}. It is essential to identify and address any disparities in participation or outcomes among different student groups, promoting inclusivity and fairness in voucher initiatives.
    \item \textbf{Curriculum Suggestion}: To strengthen the alignment between cybersecurity education and industry needs, schools and institutions should explore the integration of more industry-endorsed programs or certifications into their existing curricula. By offering a broader range of industry-relevant certifications, schools can better cater to the diverse career aspirations of students within the cybersecurity field \cite{mardis2018assessing}. This approach fosters a curriculum that is responsive to industry trends and demands.
    \item \textbf{Training}: Members involved in cybersecurity education should undergo comprehensive training tailored to the effective management of voucher incentive programs. This training should equip educators with the skills and knowledge needed to address the challenges and opportunities associated with such programs \cite{toth2013role}. This training should encompass areas such as academic integrity, assessment design, and strategies for providing robust support to students pursuing voucher incentives. Empowering educators with the necessary tools ensures the successful implementation of voucher initiatives and supports student success.
    \item \textbf{Resource Allocation}: To facilitate the achievement of High Distinctions by motivated students, schools and institutions should allocate resources specifically aimed at supporting these students. This support includes providing additional academic assistance, granting access to relevant study materials, and offering exam preparation assistance \cite{knapp2017maintaining}. By ensuring that students aiming for High Distinctions have access to the necessary resources, institutions maximize their chances of success and create an environment conducive to academic excellence.
\end{itemize}

\subsection{Voucher Selection}
When integrating industry vouchers into CSE curricula, it is important to select vouchers that align with educational objectives and students' employability. Here's a suggested process:

\begin{enumerate}
    \item \textbf{Identify relevant industry certifications}: Research industry-standard certifications relevant to cybersecurity education, such as CompTIA Security+, Certified Information Systems Security Professional (CISSP), Certified Ethical Hacker (CEH), and Certified Secure Computer User (CSCU), etc. Look for certifications that are recognized by employers and are known to enhance students' employability.
    \item \textbf{Analyze job market trends}: Study current job market trends in cybersecurity to identify in-demand skills and certifications. Look for certifications that are frequently mentioned in job postings and are preferred by employers in the field.
    \item \textbf{Consult with industry professionals}: Seek input from industry professionals, employers, and alumni who are actively working in the cybersecurity field. %Ask for their insights on which certifications are most valued and sought after by employers.
    \item \textbf{Consider regional preferences}: Take into account regional variations in certification preferences and requirements. %Some certifications may be more relevant or in demand in certain geographic areas due to industry-specific needs or regulatory requirements.
    \item  \textbf{Assess curriculum alignment}: Assess the current cybersecurity curriculum to identify intersections with industry certifications and course objectives. Determine which certifications most effectively complement the skills and knowledge intended to be imparted to students. %Aligning education with industry needs ensures that all students can derive value from voucher incentives, regardless of whether they ultimately receive a voucher.
    \item \textbf{Monitor industry updates}: Stay informed about changes and updates in the cybersecurity industry, including emerging certifications and evolving skill requirements. %Continuously reassess and update the selection of vouchers to ensure they remain relevant and beneficial for students' employability.
\end{enumerate}

By following these steps, educators can effectively select industry vouchers and integrate them into CSE curricula to enhance students' learning and career opportunities.

%如果你觉得太长了，可以删减。比如5， 7， and 8 可以简单提一下？不用单独列出来？

\subsection{Limitations}

While this mixed-methods case study approach offers a comprehensive perspective on the research phenomenon, it is essential to acknowledge certain limitations. The study's findings may be context-specific to the university and the investigated unit, limiting generalizability. Furthermore, the research design is retrospective, capturing students' perceptions after completing their units, which may introduce recall bias \cite{blome2015measuring}.

\section{Conclusion}
\label{sec: conclusion}
In this paper, we investigated the motivational potential of voucher incentives within cybersecurity education (CSE). Using a cybersecurity foundation unit as a case study, we examined the impact of industry vouchers on students’ academic aspirations, engagement, and perceptions of effectiveness.

First and foremost, voucher incentives have proven to be powerful motivators, significantly influencing students’ academic goal-setting. These incentives fostered increased engagement with course materials and prompted students to cultivate more effective study habits, underscoring their positive impact on the learning experience. The positive student perceptions of voucher effectiveness highlight their value as tools for educational enhancement.

Moreover, we recommend future research endeavours to explore the long-term impact of voucher incentives on students’ career trajectories and employability. In the broader context of educational research, this study underscores the significance of methodological innovation and approaches to improve learning outcomes. 
Voucher incentives offer a promising avenue for exploration and innovation in educational
settings.

Overall, our research has elucidated the potential of voucher incentives to transform CSE, motivating students to pursue academic excellence and fostering deeper engagement. As educational institutions and researchers continue to seek innovative strategies to enhance education, voucher incentives stand out as a compelling tool with the potential to make a lasting impact.

%We have provided valuable insights into the transformative potential of voucher incentives within cybersecurity education (CSE). 
%, providing a comprehensive understanding of their role in shaping future cybersecurity professionals

%Voucher recipients displayed a heightened commitment to achieving high distinctions, reflecting a strong desire to excel academically. 
%Students recognized vouchers as valuable resources that reward academic excellence and encourage sustained effort and commitment.

%\input{referenc}
\bibliographystyle{unsrt}
\bibliography{csce.bib}
\end{document}